\RequirePackage{lineno}   % for line numbers
\documentclass[prd,twocolumn,showpacs,amsmath,amssymb,superscriptaddress]{revtex4}

\usepackage{graphicx}% Include figure files
\usepackage{dcolumn}% Align table columns on decimal point
\usepackage{bm}% bold math

\providecommand{\e}[1]{\ensuremath{\times 10^{#1}}}
\providecommand{\fluxunits}{\mathrm{cm^{-2} sr^{-1} s^{-1}}}

\begin{document}

\title{Search for Relativistic Magnetic Monopoles with IceCube}

\affiliation{III. Physikalisches Institut, RWTH Aachen University, D-52056 Aachen, Germany}
\affiliation{School of Chemistry \& Physics, University of Adelaide, Adelaide SA, 5005 Australia}
\affiliation{Dept.~of Physics and Astronomy, University of Alaska Anchorage, 3211 Providence Dr., Anchorage, AK 99508, USA}
\affiliation{CTSPS, Clark-Atlanta University, Atlanta, GA 30314, USA}
\affiliation{School of Physics and Center for Relativistic Astrophysics, Georgia Institute of Technology, Atlanta, GA 30332, USA}
\affiliation{Dept.~of Physics, Southern University, Baton Rouge, LA 70813, USA}
\affiliation{Dept.~of Physics, University of California, Berkeley, CA 94720, USA}
\affiliation{Lawrence Berkeley National Laboratory, Berkeley, CA 94720, USA}
\affiliation{Institut f\"ur Physik, Humboldt-Universit\"at zu Berlin, D-12489 Berlin, Germany}
\affiliation{Fakult\"at f\"ur Physik \& Astronomie, Ruhr-Universit\"at Bochum, D-44780 Bochum, Germany}
\affiliation{Physikalisches Institut, Universit\"at Bonn, Nussallee 12, D-53115 Bonn, Germany}
\affiliation{Dept.~of Physics, University of the West Indies, Cave Hill Campus, Bridgetown BB11000, Barbados}
\affiliation{Universit\'e Libre de Bruxelles, Science Faculty CP230, B-1050 Brussels, Belgium}
\affiliation{Vrije Universiteit Brussel, Dienst ELEM, B-1050 Brussels, Belgium}
\affiliation{Dept.~of Physics, Chiba University, Chiba 263-8522, Japan}
\affiliation{Dept.~of Physics and Astronomy, University of Canterbury, Private Bag 4800, Christchurch, New Zealand}
\affiliation{Dept.~of Physics, University of Maryland, College Park, MD 20742, USA}
\affiliation{Dept.~of Physics and Center for Cosmology and Astro-Particle Physics, Ohio State University, Columbus, OH 43210, USA}
\affiliation{Dept.~of Astronomy, Ohio State University, Columbus, OH 43210, USA}
\affiliation{Dept.~of Physics, TU Dortmund University, D-44221 Dortmund, Germany}
\affiliation{Dept.~of Physics, University of Alberta, Edmonton, Alberta, Canada T6G 2G7}
\affiliation{D\'epartement de physique nucl\'eaire et corpusculaire, Universit\'e de Gen\`eve, CH-1211 Gen\`eve, Switzerland}
\affiliation{Dept.~of Physics and Astronomy, University of Gent, B-9000 Gent, Belgium}
\affiliation{Dept.~of Physics and Astronomy, University of California, Irvine, CA 92697, USA}
\affiliation{Laboratory for High Energy Physics, \'Ecole Polytechnique F\'ed\'erale, CH-1015 Lausanne, Switzerland}
\affiliation{Dept.~of Physics and Astronomy, University of Kansas, Lawrence, KS 66045, USA}
\affiliation{Dept.~of Astronomy, University of Wisconsin, Madison, WI 53706, USA}
\affiliation{Dept.~of Physics and Wisconsin IceCube Particle Astrophysics Center, University of Wisconsin, Madison, WI 53706, USA}
\affiliation{Institute of Physics, University of Mainz, Staudinger Weg 7, D-55099 Mainz, Germany}
\affiliation{Universit\'e de Mons, 7000 Mons, Belgium}
\affiliation{T.U. Munich, D-85748 Garching, Germany}
\affiliation{Bartol Research Institute and Department of Physics and Astronomy, University of Delaware, Newark, DE 19716, USA}
\affiliation{Dept.~of Physics, University of Oxford, 1 Keble Road, Oxford OX1 3NP, UK}
\affiliation{Dept.~of Physics, University of Wisconsin, River Falls, WI 54022, USA}
\affiliation{Oskar Klein Centre and Dept.~of Physics, Stockholm University, SE-10691 Stockholm, Sweden}
\affiliation{Department of Physics and Astronomy, Stony Brook University, Stony Brook, NY 11794-3800, USA}
\affiliation{Dept.~of Physics and Astronomy, University of Alabama, Tuscaloosa, AL 35487, USA}
\affiliation{Dept.~of Astronomy and Astrophysics, Pennsylvania State University, University Park, PA 16802, USA}
\affiliation{Dept.~of Physics, Pennsylvania State University, University Park, PA 16802, USA}
\affiliation{Dept.~of Physics and Astronomy, Uppsala University, Box 516, S-75120 Uppsala, Sweden}
\affiliation{Dept.~of Physics, University of Wuppertal, D-42119 Wuppertal, Germany}
\affiliation{DESY, D-15735 Zeuthen, Germany}

\author{R.~Abbasi}
\affiliation{Dept.~of Physics and Wisconsin IceCube Particle Astrophysics Center, University of Wisconsin, Madison, WI 53706, USA}
\author{Y.~Abdou}
\affiliation{Dept.~of Physics and Astronomy, University of Gent, B-9000 Gent, Belgium}
\author{M.~Ackermann}
\affiliation{DESY, D-15735 Zeuthen, Germany}
\author{J.~Adams}
\affiliation{Dept.~of Physics and Astronomy, University of Canterbury, Private Bag 4800, Christchurch, New Zealand}
\author{J.~A.~Aguilar}
\affiliation{D\'epartement de physique nucl\'eaire et corpusculaire, Universit\'e de Gen\`eve, CH-1211 Gen\`eve, Switzerland}
\author{M.~Ahlers}
\affiliation{Dept.~of Physics and Wisconsin IceCube Particle Astrophysics Center, University of Wisconsin, Madison, WI 53706, USA}
\author{D.~Altmann}
\affiliation{Institut f\"ur Physik, Humboldt-Universit\"at zu Berlin, D-12489 Berlin, Germany}
\author{K.~Andeen}
\affiliation{Dept.~of Physics and Wisconsin IceCube Particle Astrophysics Center, University of Wisconsin, Madison, WI 53706, USA}
\author{J.~Auffenberg}
\affiliation{Dept.~of Physics and Wisconsin IceCube Particle Astrophysics Center, University of Wisconsin, Madison, WI 53706, USA}
\author{X.~Bai}
\thanks{Physics Department, South Dakota School of Mines and Technology, Rapid City, SD 57701, USA}
\affiliation{Bartol Research Institute and Department of Physics and Astronomy, University of Delaware, Newark, DE 19716, USA}
% ...

\collaboration{IceCube Collaboration}
\noaffiliation

\date{\today}

\begin{abstract}
We present the first results in the search for relativistic magnetic monopoles
with the IceCube detector, a subsurface neutrino telescope located in the
South Polar ice cap containing a volume of 1 km$^{3}$.  
This analysis searches data taken on the partially
completed detector during 2007 when roughly 0.2 km$^{3}$ of ice
was instrumented.  The lack of candidate events
leads to an upper limit on the flux of relativistic magnetic monopoles of
$\Phi_{\mathrm{90\%C.L.}}\sim 3\e{-18}\fluxunits$
for $\beta\geq0.8$. 
This is a factor of 4 improvement over 
the previous best experimental flux limits up to a Lorentz boost $\gamma$ 
below $10^{7}$.
This result is then interpreted for a wide range
of mass and kinetic energy values.
\end{abstract}

\pacs{14.80.Hv} 

\maketitle
%% \linenumbers

\newpage
\section{\label{sec:Intro}Introduction}
Magnetic monopoles are an important element in a complete picture
of our universe.  Their existence would explain the quantization
of electric (and magnetic) charge via the Dirac quantization
equation $g=Ne/2\alpha$ \cite{Dirac}.  They appear
as topological defects from symmetry breaking in Grand Unified Theories
(GUTs) \cite{tHooftPoly} with
masses $\sim10^{4}-10^{17}\,$GeV \cite{Wick}, depending on the breaking scheme. 
Additionally, they would bring a complete symmetry to Maxwell's equations.

Magnetic monopoles produced in the early universe via GUT symmetry breaking
would be topologically stable and accelerated along 
magnetic field lines.  The universe is full of long range magnetic fields
that would act upon the monopoles over their lifetime, likely imparting
energies $\sim10^{15}\,$GeV \cite{Wick}.  Therefore,
magnetic monopoles below this energy scale should 
reach and travel through the Earth
at relativistic speeds.   
A relativistic magnetic monopole 
moving through a transparent medium would produce copious amounts of Cherenkov 
light, $\sim$8300 times a single muon in ice \cite{Tom}.  
Thus, large Cherenkov detectors like IceCube are an ideal 
experiment to search for these particles.

The current best limits on the flux of magnetic monopoles 
at the 90\% confidence level (C.L.) for relativistic speeds 
between $\beta=0.8$ and Lorentz boost
$\gamma=10^{7}$ are set by the ANTARES detector \cite{ANTARES} at
the $\sim 10^{-17}\fluxunits$ scale.  This recent result is the first
in this velocity range to 
surpass the results from the AMANDA detector \cite{AMANDA}, IceCube's proof of concept, which
set flux limits $\sim3\e{-17}\fluxunits$.  ANTARES also searched
for magnetic monopoles below the Cherenkov threshold but
still energetic enough to knock off electrons that produce Cherenkov light.  This
extension sets flux limits at the $\sim5\e{-17}\fluxunits$ scale down
to a speed of $\beta=0.625$. 
For lower speeds, MACRO provides comprehensive
flux limits $\sim10^{-16}\fluxunits$ \cite{MACRO} for speeds down to $\beta\sim10^{-4}$
while flux limits at
ultra-relativistic speeds are set by radio detectors RICE \cite{RICE} and
ANITA \cite{ANITA} at the $\sim10^{-19}\fluxunits$ scale.  

These are important
as they are flux limits below the 'Parker Bound' \cite{PB} 
($\sim 10^{-15}\fluxunits$), an astrophysical
flux limit derived by considering the survival of the galactic magnetic field
in the presence of magnetic monopoles. More sophisticated calculations that consider
velocity \cite{Turner} relax the bound on relativistic magnetic monopoles above
a mass of $10^{11}\,$GeV due to the shortened time spent in the galactic field.
However, an 'Extended Parker Bound'
found by considering the survival of a modeled seed field still
produces flux limits
well below experiments, with 
$\Phi\sim10^{-16}(\mathrm{Mass_{MP}})/(10^{17}\,\mathrm{GeV})\,\fluxunits$ 
\cite{EPB}.

This paper describes the search for relativistic magnetic monopoles in
data taken with the IceCube detector between May 2007 and April 2008.
The analysis is optimized for magnetic monopoles with modest Lorentz
boosts ($\gamma\le10$) and charge $g=1$.  
The derived flux limits are conservative upper bounds for magnetic
monopoles with larger $\gamma$ or charge, as these cases produce more
light in the ice.
The paper is organized as follows.  Section \ref{sec:Det} describes the IceCube 
detector.  Section \ref{sec:Sim} describes the simulation of background and
signal. Section \ref{sec:EvSel} defines the variables and outlines the
steps used to discriminate signal events from background.  
Section \ref{sec:Unc} summarizes the uncertainties.  
Section \ref{sec:Res} presents the results for an isotropic flux of
magnetic monopoles at the detector.  Section \ref{sec:Dis} extends this
result to an isotropic flux at the Earth's surface by considering the
energy loss
of magnetic monopoles through the Earth.  
% These results are presented
% as a function of both the magnetic monopole mass and kinetic energy, allowing
% for the broad range of uncertainty in each of these values.  
This results in a final limit plot that is presented over a large
range of magnetic mass and kinetic energy values.  This 
allows the result to remain agnostic towards the particular 
origin and energy gaining mechanism a magnetic monopole 
may possess.  Concluding remarks
are presented in Section \ref{sec:Conc}.

\section{\label{sec:Det}IceCube Detector}

\begin{figure*}
\includegraphics[width=.32\textwidth]{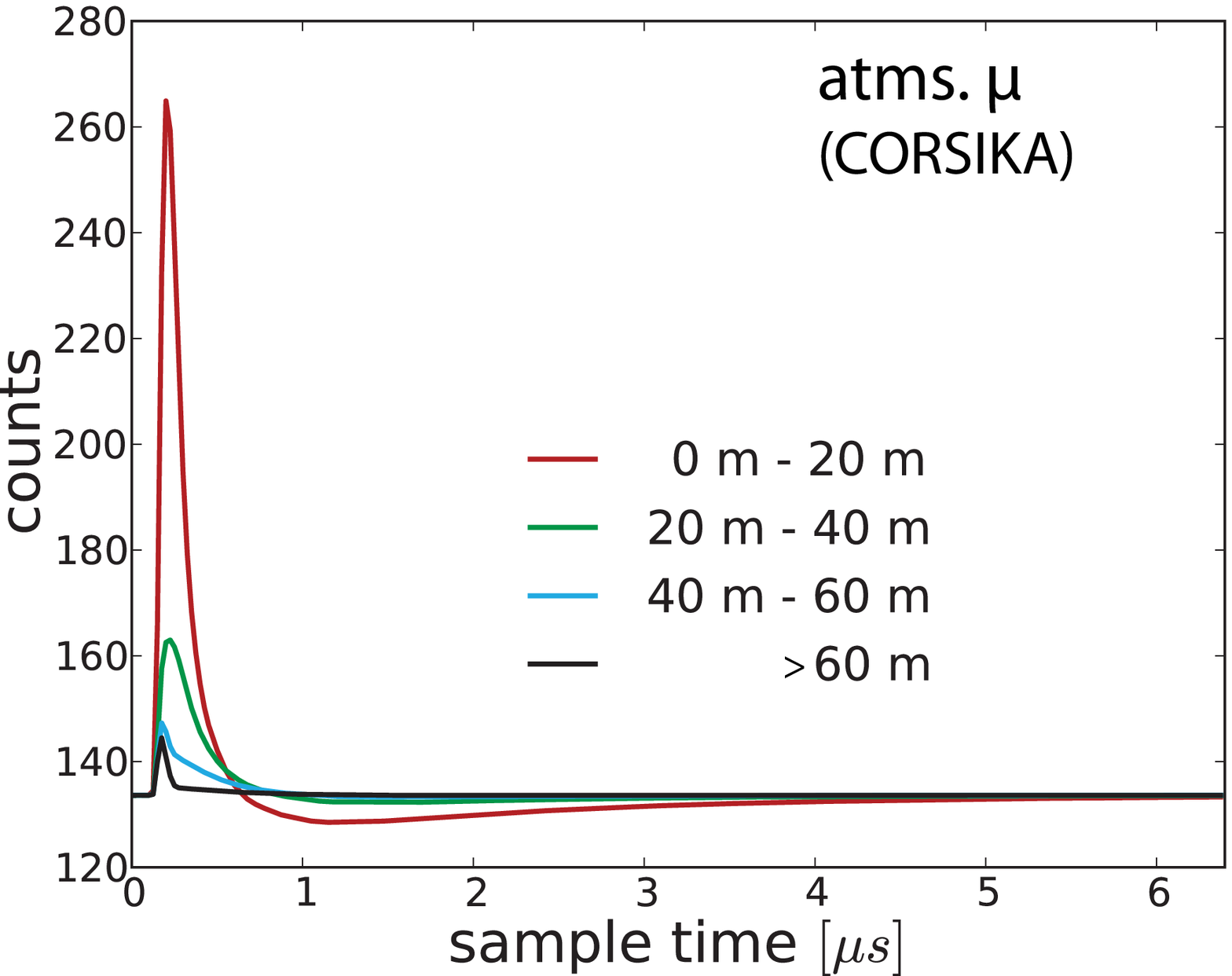}
\includegraphics[width=.32\textwidth]{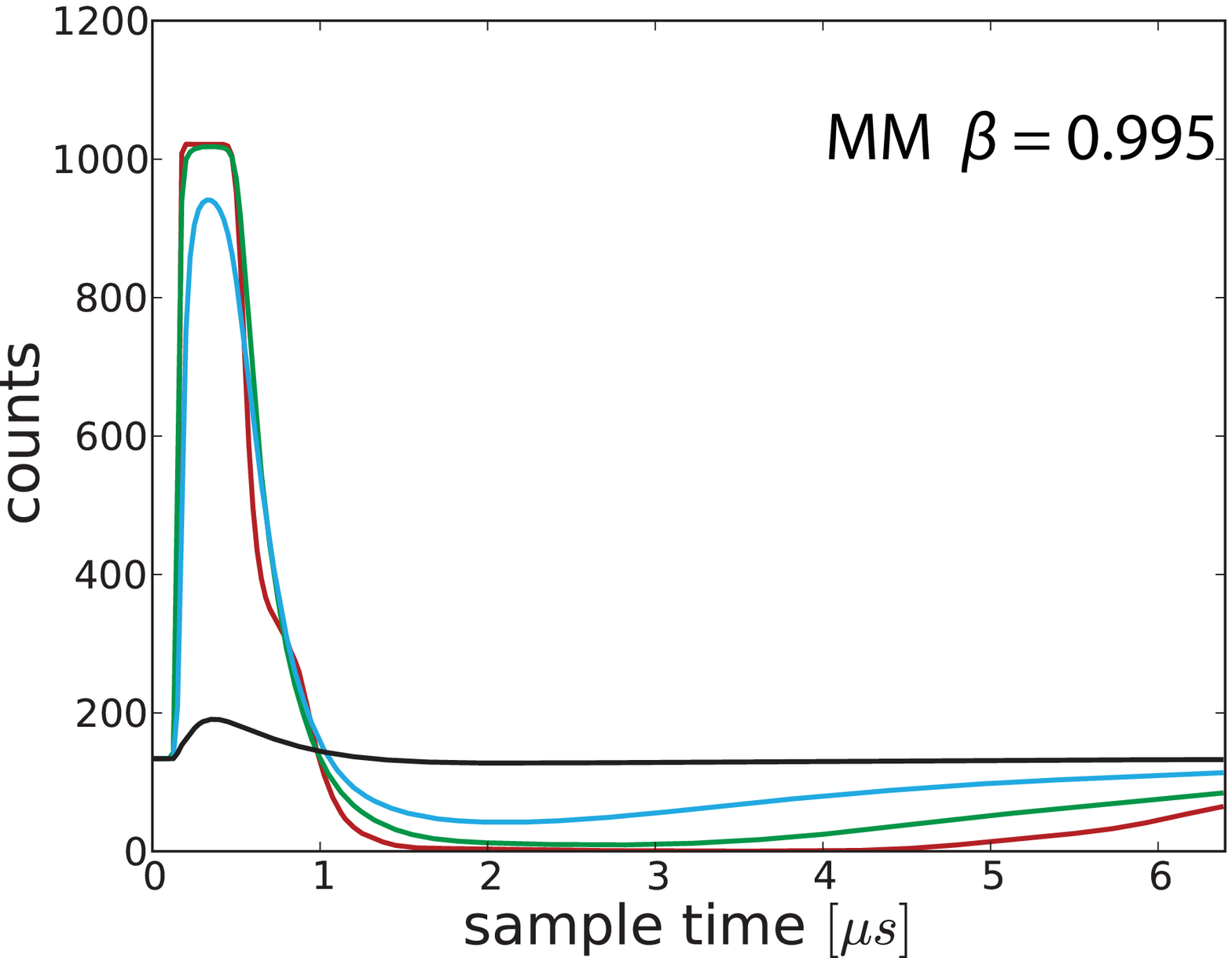}
\includegraphics[width=.32\textwidth]{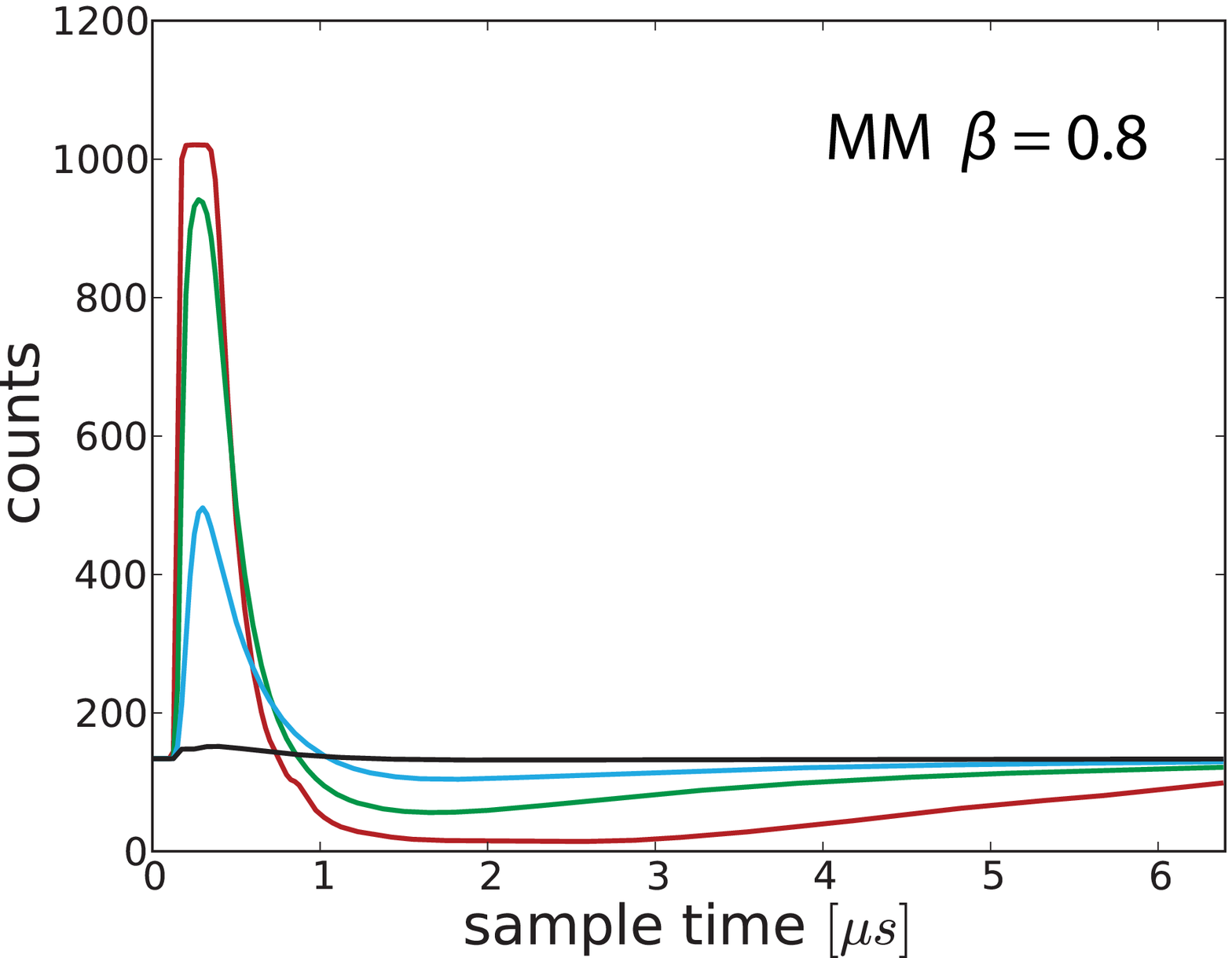}
\caption{Averaged PMT ADC waveforms for ~1000 simulated events of atmospheric muon
  background (left) as well as the $\beta=0.995$ (middle) and $\beta=0.8$ (right) magnetic
  monopoles.  Distances are how far away the particle is from
  the DOM receiving the light.}
\label{fig:WFs} 
\end{figure*}

IceCube is a telescope at the South Pole which 
detects neutrinos by measuring the Cherenkov light from secondary
charged particles produced in neutrino-nucleon interactions \cite{IceCube}.
A total of 5160 Digital Optical Modules (DOMs)
are arranged in 86 vertical strings frozen
in the ice between 1500 and 2500$\,$m below the surface over a 
total volume of 1 $\mathrm{km}^{3}$.  Construction
was completed in December 2010.  The data for this analysis were
taken during the construction phase, when only 22 of the 86 strings
had been deployed.  The 22 strings contain a volume of 
$\sim0.2\,\mathrm{km}^{3}$.

The DOM is the centerpiece of the IceCube detector and houses a 10-inch 
photomultiplier tube (PMT) to detect light, 
onboard electronics for pulse digitization,
and LED light sources for calibration.  
Light signals which pass a threshold of 0.25 photo-electron (PE) 
PMT pulse heights are digitized and the DOM is said to 'launch'.  
Two types of waveform digitizers are utilized.  The
Analog Transient Waveform Digitizer (ATWD) bins the waveform with a 3.3$\,$ns 
sampling period over a readout window of 420$\,$ns.  It supports three
channels with different gains in order to extend its effective dynamic range.
The PMT Analog to Digital Converter 
(ADC) collects data at a slower sampling rate of 25$\,$ns and records
for 6.4$\,\mu\mathrm{s}$.  

A time is calculated for the launch by re-syncing the threshold
crossing to the next leading edge of the internal DOM clock, which
oscillates at 40 MHz.  More precise timing is achieved by later
reconstructing the leading edge of the digitized waveform, though in
this analysis the coarse time is sufficient.
For more information on the DOM and its components, see \cite{DOM,PMT}.

Each waveform digitizer outputs the signal in terms 
of counts/bin values that directly map to the voltage recorded. For the PMT
ADC, which is the only channel used in this analysis, 
a single photo-electron corresponds to 
$\sim13$ counts \cite{DOM}.  
The PMT ADC saturates at 1024 counts, which occurs when
$\sim50-100\,$PE's arrive in a single 25$\,$ns bin.  
More generally, DOM's that receive $\sim600\,$PE's over the full readout window 
typically saturate.
Figure \ref{fig:WFs} shows example PMT ADC waveforms for both
background and signal events at various
distances to the DOM.  The flattened top for the signal indicates the point where
the digitizer saturates.

Once a launch is recorded, the DOM checks the four nearest neighbors on the string
to see if another hit occurred within a 1$\,\mathrm{\mu}$s time window.  
By requiring companion launches to occur, the effect of dark noise
hits is reduced.  If this local
coincidence condition is met, the digitized waveforms and time are sent to 
the surface.  A trigger
algorithm is applied to determine if a physics event has been detected.  For the
22 string detector, this algorithm checked if 8 hits were recorded within a sliding
5$\,\mu\mathrm{s}$ time window.  
For data in this analysis, the average trigger rate is $\sim550\,$Hz and is vastly
dominated by muons generated in cosmic ray air showers in the
atmosphere above the South Pole. 

Against this background a magnetic 
monopole event would stand out due to the much 
higher light deposition. For this analysis a further filter is applied
to the data online at the South Pole, requiring at least 80 DOM
launches in an event \cite{EHEPaper}.  This retains all bright events,
regardless of direction.  The passing rate for this filter is 
$\sim1.5\,$Hz.  It consists of muon bundles containing hundreds of muons
generated by high energy cosmic ray primaries.  
All data that pass this filter are sent north via satellite to a data
warehouse for use by the entire collaboration.

\section{\label{sec:Sim}Simulation of Datasets}
Simulation of the background and signal are done within the \texttt{ICETRAY}
framework, a \texttt{C++} based code written for use by the IceCube 
Collaboration.  It includes tools to simulate the detector response to
light produced by particles as well as the triggering and filtering
algorithms.  This allows simulated events to be compared directly to the
experimental data.

\subsection{\label{sec:BSim}Background Datasets}
Background simulation is composed of muon bundles and neutrinos produced in the 
atmosphere by high energy cosmic rays.  The generation uses importance sampling
in energy so that at the final analysis level the statistical uncertainty in 
background prediction is of the order of systematic uncertainty or less.

Atmospheric muon bundles are
simulated with \texttt{CORSIKA} \cite{Cors} using two primary types: proton
to represent light elements and iron to represent heavier ions.  Primary energies
are simulated between $10^{4}$ and $10^{11}\,$GeV.  Events are generated
with an $E^{-2}$ spectrum to oversample the high energy region.
The events are weighted to
fits of extensive air showers introduced by the KASCADE Collaboration
\cite{2comp}. The muon bundles are then propagated through the ice
using \texttt{MMC}(Muon Monte Carlo) \cite{MMC}.

\texttt{ANIS}(All Neutrino Interaction Simulation) \cite{ANIS} 
is used to simulate both muon and electron neutrino events.
The neutrinos are generated with an $E^{-1}$ spectrum and given weights
corresponding to a conventional atmospheric neutrino flux from Honda \cite{Honda} 
and a prompt flux from charmed meson production based on the 
Enberg, Reno, and Sarcevic model \cite{Sarc}.

\subsection{\label{sec:SSim}Signal Datasets}
Code developed specifically for this analysis 
is used to generate and propagate the signal magnetic monopoles.
Three datasets are created for discrete speeds of $\beta=0.8, 0.9$ and 
$0.995\,(\gamma=10)$.
Monopole tracks are generated by randomly distributing vertices on a circular ``generation plane'' % ($A_{\rm gen}$) 
with radius 650\,m at a distance of 1000\,m from the detector center. From the vertices, monopoles are propagated towards 
and through the detector with directions perpendicular to the plane. During generation, the orientations of the 
generation plane relative to the detector are randomized, thereby creating an isotropic monopole flux through the detector.

Above $\beta\sim0.1$ and below $\gamma\sim10^{4}$,
the electromagnetic energy loss of magnetic monopoles
through matter is well described by a combination of ionization and atomic
excitations, collectively referred to as 'collisional' energy loss \cite{Ahlen}.
As the choice of simulated events only reach $\gamma=10$, this is the only
energy loss considered in propagation.  Above $\gamma\sim10^{4}$, energy losses
from pair production and photo-nuclear interactions surpass
the collisional losses.
These energy losses are considered in
Section \ref{sec:AngAccEarth} for magnetic monopoles traveling through the Earth
with large boost factors.  Bremsstrahlung, which
is proportional to $1/M^{2}$, is heavily suppressed.

For each dataset, 100,000 events are generated at a mass of $M=10^{11}\,$GeV.  
The effect of choosing one mass is mitigated since the Cherenkov light output
only depends on speed which remains essentially constant over the 1.2$\,$km 
path through the detector.

\section{\label{sec:EvSel} Event Selection}

\begin{figure}
\includegraphics[scale=0.4]{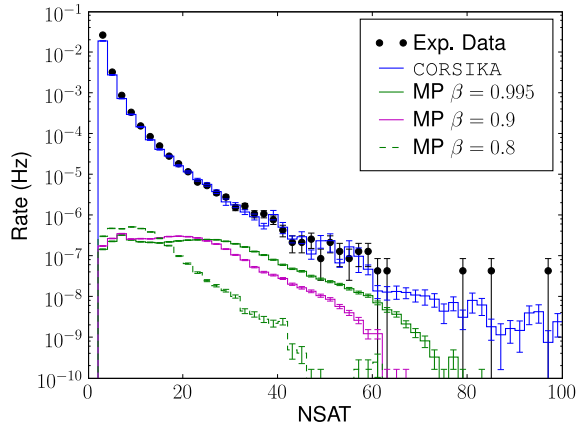}
\caption{The number of saturated hits per event for the simulated signal and 
atmospheric muon background (\texttt{CORSIKA}).  In addition, the full experimental
data set is included.}
\label{fig:NSat} 
\end{figure}

The main strategy employed to select relativistic magnetic monopoles 
is to look for extremely bright events.  
This is measured by counting the number of DOM launches which capture
a high charge.  High charge DOM launches are defined as ones that 
saturate the PMT ADC channel.
Figure \ref{fig:NSat} shows the number of these ``saturated hits''
(NSAT).  To visualize
the signal event rates, a flux of $5\e{-17}\fluxunits$ is used.

 A secondary strategy is to exploit the arrival directions of the
incoming particle tracks.  The dominant background of 
atmospheric muon bundles can only reach the detector from above the horizon.  
This background can be suppressed by focusing on events with arrival
directions below the horizon. 

Event selection consists of three phases.  First, a simple filter is applied
to reduce the data to a manageable size.  Then, particle tracks are 
reconstructed and poorly reconstructed events are rejected 
using quality cuts.  At the final stage, an optimized cut which maximizes the Model Rejection
Factor (MRF) \cite{Hill} is found.  To reduce experimenter bias, the maximized 
MRF is found using simulated background alone.  The resulting cut is then applied
to the experimental data.  

Table \ref{tab:Rates} displays the final event rates (in events/year) for
each of the datasets considered at all levels of the analysis.  

\subsection{Track Reconstruction}

Since directional information is 
used mainly to distinguish between up and down-going particles, pointing
accuracy is only of secondary importance. Contrary to most IceCube 
analyses, which use computationally intensive likelihood methods to
reconstruct the particle tracks with sub-degree accuracy, a very 
fast analytic fit proved sufficient for this analysis.

The fundamental piece of datum used by the reconstruction is a ``hit'', which is defined as the 
location $\vec{X}$ and a time $t$ of a DOM launch. The track direction and particle speed are 
reconstructed by a least-squares fit of the observed hit pattern $\left\{ \vec{X}_{\rm i}, t_{\rm i} \right\}$ to a plane wave
of light, whose analytic solution is given by \cite{RecoPaper}

\begin{eqnarray}
\vec{X}=\vec{X}_{\rm{avg}}+\vec{V}t
\\
\vec{V}=\frac{\sum{(\vec{X}_{i}-\vec{X}_{\rm{avg}})(t_{i}-t_{\rm{avg}})}}{\sum{(t_{i}-t_{\rm{avg}})^{2}}}
\label{eq:linefit}
\end{eqnarray}

where $\vec{X}_{\rm avg}$ and $t_{\rm avg}$ are the average position
and time of all the hits.
The hit times $t_i$ correspond to the time at which the DOM records a launch.
Studies of the reconstruction accuracy demonstrated this 
to be a better definition for $t_i$ than the peak time of the PMT
pulse, likely because the launch time corresponds to the arrival time of 
those Cherenkov photons which are least delayed by scattering in the
ice. The reconstructed track direction is defined by the velocity vector $\vec{V}$.

\begin{figure}
\includegraphics[scale=0.4]{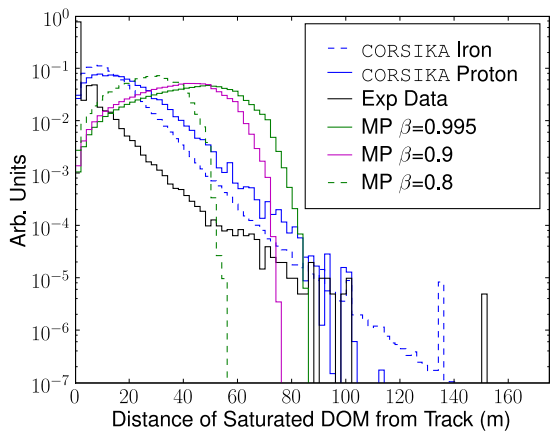}
\caption{Distance from particle track to saturated DOM.  All
  histograms are normalized to 1.  For \texttt{CORSIKA}, 
a track is defined by the primary cosmic ray.  For data, it is defined by the 
reconstructed track.  Events with only two saturated hits will reconstruct
through the hits and produce the large spike at zero.}
\label{fig:SatDist} 
\end{figure}

Because of the simple straight line hypothesis, and because the linefit does not 
take into account photon propagation through the ice, the reconstruction accuracy 
improves if only hits close to the particle track are included in the fit. 
This is achieved by selecting hits in which a large number of photons are detected.
A zenith angle resolution ($\sim 2^{\circ}$), 
defined here as the median difference between the true and 
reconstructed zenith direction for simulated events at the penultimate cut level,
 is achieved by only using hits that saturate the PMT ADC.
Shown in Fig. \ref{fig:SatDist} are the distances from the primary track to 
a saturated hit.
Saturated hits are up to $\sim10\,$m away
for muons and up to $\sim60\,$m for the fastest monopoles.
The relative closeness of the saturated hits mean timing information will be 
less affected by scattering and absorption, improving 
the accuracy of reconstructing the particle.

In addition, all hits in which the saturation occurred 
more than 500\,ns after the DOM launched are excluded from the fit. 
These are saturated hits
where the launch time is caused by something other than the saturating particle,
e.g. dark noise.  This creates errors in the reconstruction since the hit
information includes a time well before the physics event.
Roughly $0.05\%$ of the saturate hits are removed by this criterion. 

The robustness of the linefit against timing inaccuracies in the hardware 
was studied by smearing the hit times consistent with the frequency of
the internal DOM clock. This resulted in a negligible change 
on the reconstruction accuracy ($<1\%$) and final sensitivity ($<1\%$).

\subsection{Low level filter}

The low level filter selects events with high Cherenkov light yield by
requiring at least two of the hits to saturate (NSAT$>$1). 
This cut reduces background by $\sim99.5\%$ and signal by $\sim10-15\%$
and ensures that the minimum required  
two hits are available to reconstruct the track direction. 

\begin{table*}
\caption{Event rates in events/year for each dataset at all levels of the analysis.
Includes simulated signal, background and the experimental data.  For signal rates,
a flux of $5\e{-17}\fluxunits$ is assumed.}
\begin{ruledtabular}
\begin{tabular}{ccccc}
\hline 
Dataset&
Online Filter&
Low Level&
Quality Cuts&
Final\\
\hline 
Experimental Data&
$3.15\e{7}$&
$6.55\e{5}$&
$1.21\e{5}$&
$0$\\
\hline 
\hline 
\texttt{Corsika} Proton&
$7.35\e{6}$&
$2.65\e{5}$&
$2.93\e{4}$&
$3.61\e{-4}$\\
\hline 
\texttt{Corsika} Iron&
$5.14\e{6}$&
$2.20\e{5}$&
$6.19\e{4}$&
$4.70\e{-2}$\\
\hline
Atm Conv $\nu_{\mu}$&
$37.9$&
$26.4$&
$13.6$&
$3.45\e{-2}$\\
\hline
Atm Prompt $\nu_{\mu}$&
$4.9$&
$2.83$&
$0.334$&
$4.12\e{-2}$\\
\hline 
Atm Conv $\nu_{e}$&
$1.4$&
$0.967$&
$8.08\e{-6}$&
$5.52\e{-6}$\\
\hline 
Atm Prompt $\nu_{e}$&
$2.0$&
$1.86$&
$1.43\e{-3}$&
$7.39\e{-4}$\\
\hline 
\hline 
Bkgrd Total&
$1.25\e{7}$&
$4.85\e{5}$&
$9.12\e{4}$&
$0.124$\\
\hline 
\hline 
\hline 
$\beta=0.995$&
$100$&
$89.2$&
$63.4$&
$35.6$\\
\hline 
$\beta=0.9$&
$95.3$&
$84.5$&
$60.8$&
$33.4$\\
\hline 
$\beta=0.8$&
$81.0$&
$70.1$&
$46.5$&
$22.1$\\
\hline 
\end{tabular}
\end{ruledtabular}
\label{tab:Rates}
\end{table*}

\subsection{\label{sec:QualityCuts}Quality Cuts}
Background events which record several saturated hits as a result of a 
bright secondary cascade result in all the hits occurring within a small time interval and 
being located in a relatively
small volume.  This results in poor reconstructions because of the small lever arm 
to determine the overall directionality of the hits.
These are removed by requiring the saturated hits to occur over
at least 750$\,$ns.  This reduces
background by $\sim80\%$.  The signal
is reduced by $\sim30\%$, but these generally represent poor quality events that
only saturate one or two strings.

A second category of mis-reconstructed events are caused when multiple muon
bundles travel through the detector in a single trigger window.  These events
are problematic for this analysis when 
one saturates a DOM in the
bottom of the detector before a second saturates a DOM near the top, resulting
in an up-going reconstruction.  The large majority are separated enough in time
so that the speed of the reconstruction (Eq. \ref{eq:linefit}) connecting
the two events is very low.
These are eliminated by removing events with a reconstructed speed below 0.2 m/ns.
A second cut requiring the difference in $\cos{\theta}$ between the reconstruction on
all hits versus saturated hits to be within 0.6 of 0.0 eliminates these events
that happen close in time.
The combined effect of these cuts is to remove $\sim1\%$ of the
background and signal.

\subsection{\label{sec:FinalCut}Final Cut}

\begin{figure}
\includegraphics[scale=0.4]{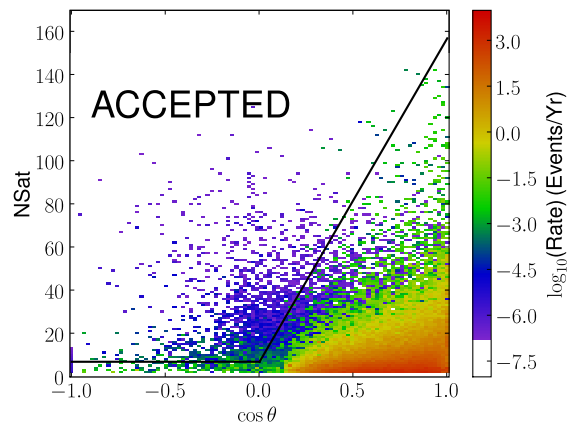}
\caption{Final cut on simulated background.  This includes atmospheric muon
bundles and atmospheric neutrinos.  Data are histogrammed with bin sizes of NSat=1 
and $\cos\theta$=0.022.}
\label{fig:2dcors} 
\end{figure}

\begin{figure}
\includegraphics[scale=0.4]{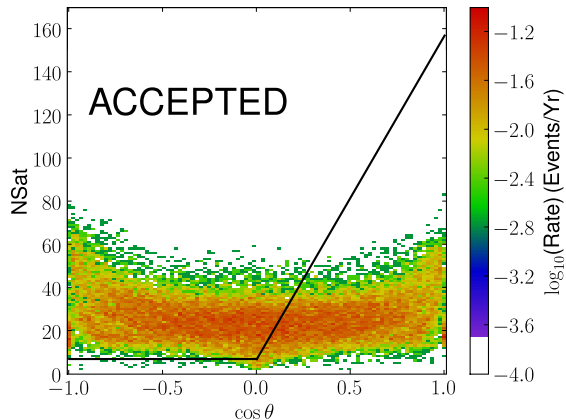}
\caption{Final cut on $\beta=0.995$ signal monopoles.  Data are histogrammed 
with bin sizes of NSat=1 and $\cos\theta$=0.022.}
\label{fig:2dsig} 
\end{figure}

The optimized final selection is a piece-wise, linear cut on NSAT and 
$\cos{\theta}$.  
Figures \ref{fig:2dcors} and \ref{fig:2dsig} show the distribution
of simulated background and the fastest monopole signal in this plane, 
along with the cut.
The background is dominated
by atmospheric muon bundles, which have a rate $\sim$5 orders of magnitude
larger than the atmospheric neutrinos.  They are essentially all down-going,
with the more vertical events
producing more saturated hits.  
Hence, the final cut is chosen to be on NSAT with an angular
dependence: constant in the up-going region ($\cos\theta<0.0$) and 
linearly increasing in strength in the down-going region
($\cos\theta>0.0$).  These two cuts join at $\cos\theta=0.0$.  
The two numbers that describe this cut are 
the value of the NSAT cut for the up-going region ('base') and the 
linear steepness for the down-going region ('slope').  
The final cut is given by:

\begin{equation}
\mbox{NSAT}> \left
\{
\begin{array}{cc}
  \mbox{base} & \mbox{if}\,\,\cos\theta<=0 \\
  \mbox{base}+\mbox{slope}*\cos\theta & \mbox{if}\,\,\cos\theta>0
\end{array}
\right.
\end{equation}

\begin{figure}
\includegraphics[scale=0.4]{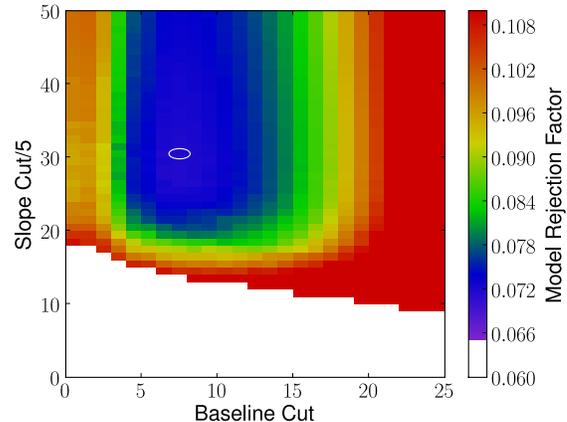}
\caption{Scan of Model Rejection Factors for final cut optimization.
  The circle corresponds to the minimum value.}
\label{fig:mrf} 
\end{figure}

A scan was made through possible values of the base from 0 to 25 in increments
of one and the slope from 0 to 250 in increments of five.  
For each possible value, the MRF \cite{Hill} is found using the event expectation
from simulation.  
Figure \ref{fig:mrf} displays the result of the scan, 
showing the stability of the minimization.  
The highlighted value corresponds to the minimum with a base of 7 and a slope of 
150.  
The final cut resulted
in a background expectation of 0.124 events/year and signal efficiencies
ranging between $\sim 47\%$ to $56\%$ relative to the penultimate cut.  

\section{\label{sec:Unc}Uncertainties}

\begin{table*}
\caption{Relative uncertainties for predicted event rates of
  background and signal.  Total uncertainties found by adding absolute
rate deviations in quadrature.}
\begin{ruledtabular}
\begin{tabular}{c|cccc|ccc}
\hline
Uncertainty&
\multicolumn{4}{c|}{Background}&
\multicolumn{3}{c}{Signal}\\
\hline
\hline
&
\texttt{CORSIKA}&
$\nu_{\mu}$&
$\nu_{e}$&
Total&
$\beta=0.8$&
$\beta=0.9$&
$\beta=0.995$\\
\hline
Normalization&
$26\%$&
$11\%$&
$<1\%$&
$12\%$&
-&
-&
-\\
\hline
Spectrum&
$990\%$&
$22\%$&
$39\%$&
$380\%$&
-&
-&
-\\
\hline
MMC cross-section&
$10\%$&
$10\%$&
-&
$7.4\%$&
-&
-&
-\\
\hline
$\nu$ cross-section&
-&
$6.4\%$&
$6.4\%$&
$4.0\%$&
-&
-&
-\\
\hline
DOM Efficiency&
$27\%$&
$38\%$&
$38\%$&
$25\%$&
$5.8\%$&
$4.8\%$&
$1.4\%$\\
\hline
Ice Properties&
$78\%$&
$40\%$&
$71\%$&
$40\%$&
$7.1\%$&
$4.2\%$&
$0.2\%$\\
\hline
Statistical&
$22\%$&
$14\%$&
$19\%$&
$12\%$&
$0.9\%$&
$0.7\%$&
$0.7\%$\\
\hline
TOTAL&
${\bf 990\%}$&
${\bf 64\%}$&
${\bf 110\%}$&
${\bf 382\%}$&
${\bf 9.2\%}$&
${\bf 6.5\%}$&
${\bf 1.7\%}$\\
\end{tabular}
\end{ruledtabular}
\label{tab:Uncertainties}
\end{table*}

Uncertainties were studied largely with Monte Carlo simulations.  Table
\ref{tab:Uncertainties} contains the results.  The large relative
background uncertainty is acceptable given the small absolute event
rate.
Uncertainties consisted of three types: (1) Theoretical
uncertainties in the simulated models, (2) Uncertainties in the detector response,
and (3) Statistical uncertainties.  

Theoretical uncertainties include the shape and normalization of the background
energy spectrum for both the atmospheric muons and neutrinos.  In addition, the
cross section uncertainty modeled in both \texttt{MMC} and \texttt{ANIS}
is studied.  Detector uncertainties include uncertainties in the scattering and 
absorption parameters of the ice and the efficiency of the DOM.  

For atmospheric muon background, the dominant
uncertainty is from the cosmic ray energy spectrum.  
For both elements, the parameters of the assumed broken power-law
(break energy, power-law indices below and above the break, and
absolute normalization) were varied within the uncertainties in the 2-component 
model \cite{2comp}. 
For iron, the extreme case of no break is taken as the 
upper end of the uncertainty since the expected break occurs beyond
the fit region of the model.  Since the final \texttt{CORSIKA} sample is 
overwhelmingly high energy iron primaries above $10^{10}\,$GeV, it is very 
sensitive to changes in the spectral weighting values.

The conservative nature of this assumption is to allow for uncertainties at the 
high energy range that are not easily tested by simulation.
Despite this extreme, the absolute uncertainty is still less than
0.5 events/year.  Signal is more robust due to the brighter light
yield.  This allows a larger sample to pass the final cuts relative to
background causing it to be less sensitive to variations in the detector response.

\section{\label{sec:Res}Results}
\begin{figure}
\includegraphics[scale=0.4]{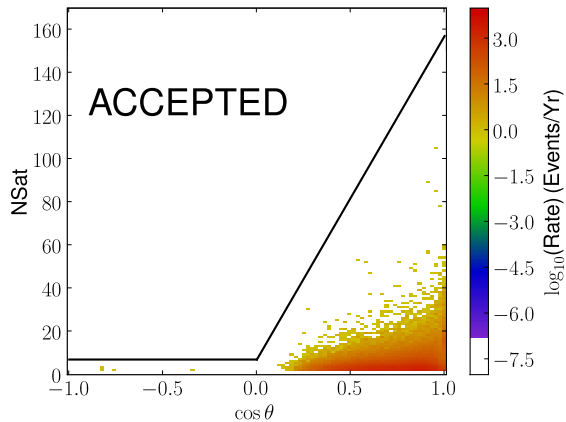}
\caption{Final cut on the full experimental data sample.  Data is histogrammed 
with bin sizes of NSat=1 and $\cos\theta$=0.022.}
\label{fig:2ddata} 
\end{figure}

The optimized cut is then applied to the full experimental data sample.  No
events survived on an expected background of 0.124 events, 
resulting in a Feldman and Cousins upper limit of 2.44 
at the 90\% C.L. \cite{FeldCous}.
The final distribution is shown in Fig.
\ref{fig:2ddata}.

The final flux limit is calculated incorporating the systematic and statistical
uncertainties using the profile log-likelihood method implemented in the
\texttt{POLE++} program \cite{Pole}.  

\begin{figure}
\includegraphics[scale=0.32]{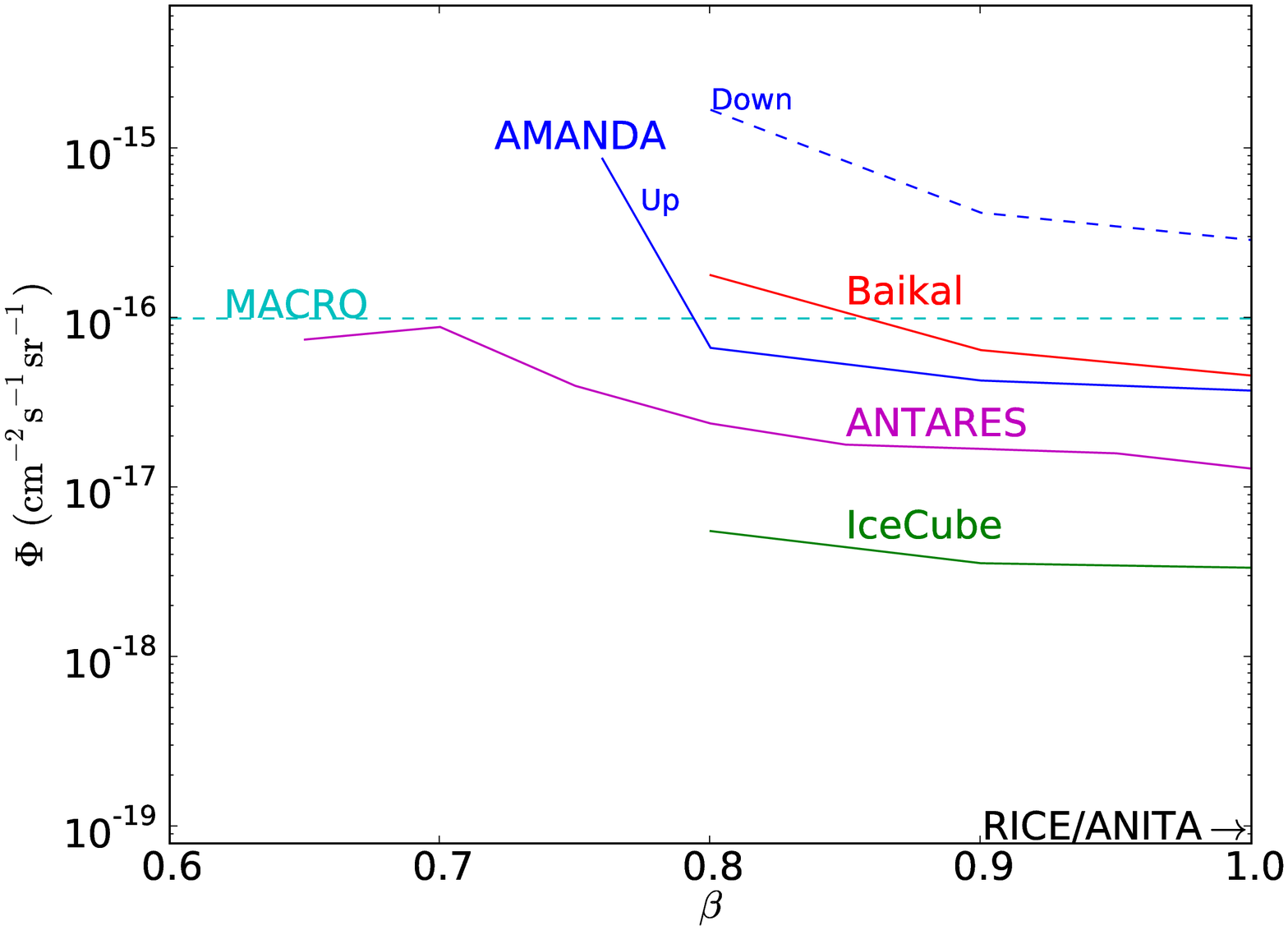}
\caption{Flux limits (90\% C.L.) at the detector as a function of $\beta$.  
Includes results from
AMANDA \cite{AMANDA}, Baikal \cite{Baikal}, ANTARES \cite{ANTARES}, 
MACRO \cite{MACRO}, RICE \cite{RICE}, and ANITA \cite{ANITA}}
\label{fig:FinalLimits_Beta} 
\end{figure}

Table \ref{tab:FinalLimits} displays
the resulting sensitivities and final limits on the flux of magnetic
monopoles at the detector at
the 90\% C.L.  Figure 
\ref{fig:FinalLimits_Beta} shows this result compared with previous searches
from neutrino telescopes.

\begin{table}
\caption{Final sensitivities and limits (90\% C.L.) on the flux of magnetic
  monopoles at detector in $\fluxunits$}
\begin{ruledtabular}
\begin{tabular}{cccc}
\hline
&
$\beta=0.8$&
$\beta=0.9$&
$\beta=0.995$\\
\hline
Sensitivity&
$6.10\e{-18}$&
$3.94\e{-18}$&
$3.73\e{-18}$\\
\hline
Final Limit&
$5.57\e{-18}$&
$3.56\e{-18}$&
$3.38\e{-18}$\\
\end{tabular}
\end{ruledtabular}
\label{tab:FinalLimits}
\end{table}

\section{\label{sec:Dis}Discussion}
In order to describe the results as pertaining to an isotropic flux at the
surface of the Earth, the efficiency of the analysis as a function of zenith
is combined with the acceptance of relativistic magnetic monopoles through the Earth.  
The previous AMANDA
analysis did a similar procedure \cite{AMANDA}.  

\subsection{\label{sec:AngAccEarth}Angular Acceptance Through the Earth}
For an isotropic, mono-energetic flux
$\Phi^{\gamma_{s},M}$ of magnetic monopoles with mass M and kinetic energy 
$E_{\mathrm{Kin}}=M(\gamma_{s}-1)$ at the Earth's surface,  
the resulting $\gamma$ of the monopole flux at the detector is calculated for
$\cos{\theta}$ values in increments of 0.1.  The energy loss
is modeled using Ahlen's
stopping power formula for collisional loss \cite{Ahlen} and code
from ANITA \cite{ANITA} for pair production and photo-nuclear losses.  

\begin{figure}
\includegraphics[scale=0.32]{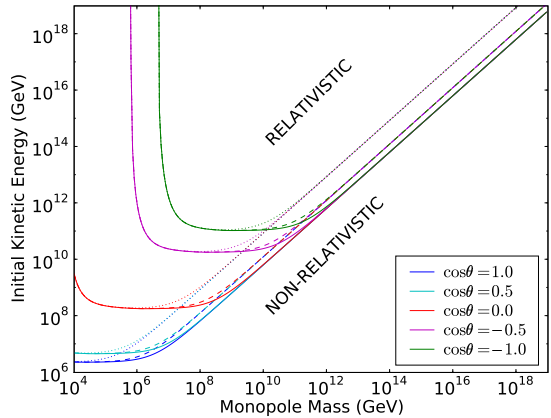}
\caption{Parameter space of magnetic monopoles at the Earth's surface
  required to reach the IceCube detector at relativistic speeds.  
Dotted is for a speed threshold of $\beta=0.995$, dashed for $\beta=0.9$, 
and solid for $\beta=0.8$.}
\label{fig:PSpace} 
\end{figure}

Figure \ref{fig:PSpace} shows the angular acceptance of relativistic magnetic 
monopoles
traveling through the Earth.  Each line indicates the boundary between
the mass and kinetic energy values which allow the monopole to reach
IceCube at a particular speed threshold for a given zenith. 
For instance, if the mass and kinetic energy are in the region 
above the $\cos\theta=-1.0$ line, these describe a magnetic monopole
that can remain relativistic traveling the diameter of the
Earth, while those above the $\cos\theta=1.0$ can remain relativistic
traveling through
the atmosphere and the $\sim 2\,$km of South Polar ice to reach the detector.

The shape of the lines can be understood
by considering the full acceptance ($\cos\theta=-1.0$) case:
\begin{list}{*}{}
\item The collisional energy loss straight up through the Earth is 
$\sim10^{11}\,$GeV.  This loss is not enough to slow relativistic
magnetic monopoles with masses above $\sim 10^{12}\,$GeV 
to sub-relativistic speeds.  Therefore, the acceptance 
is determined solely by the starting energy.

\item Magnetic monopoles with masses between $\sim
  10^{7}-10^{12}\,$GeV can still reach the detector
so long as there is enough kinetic energy to overcome the collisional loss.
Hence, the line flattens out around $\sim 10^{11}\,$GeV.

\item For magnetic monopoles with masses below $\sim 10^{7}\,$GeV, 
the necessary starting energy 
begins to increase to overcome the increasing effect of pair production 
and photo-nuclear energy
losses, which begin to dominate for $\gamma\sim 10^{4}$.  
\end{list}

\subsection{\label{sec:AngAccAnalysis}Angular Acceptance of Analysis}
The analysis
is much more sensitive to an up-going signal, due to the large
atmospheric muon bundle background.  This is described quantitatively 
by calculating the
effective area as a function of zenith.  
The effective area corresponds to
the cross sectional area of an ideal detector with 100\% efficiency.
Using the same $\cos{\theta}$ bins
as above, the effective area is given by:
\begin{equation}
A^{\gamma}_{\rm{eff}}(\cos{\theta})=A^{\gamma}_{\rm{gen}}\frac{N_{\rm{det}}^{\gamma}(\cos{\theta})}{N_{\rm{gen}}^{\gamma}(cos{\theta})}
\end{equation}
where $A_{\rm{gen}}^{\gamma}$ is the area of the 
generation plane for a given $\gamma$
and $N_{\rm{det}}^{\gamma}/N_{\rm{gen}}^{\gamma}$ is the 
fraction of magnetic monopoles generated that survive the final analysis cut.

\begin{figure}
\includegraphics[scale=0.4]{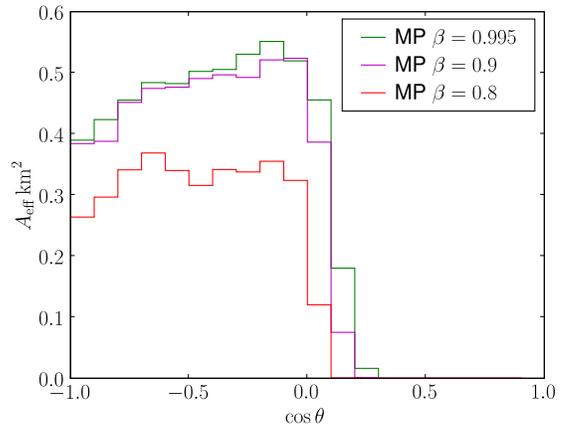}
\caption{Effective Area for each $\cos{\theta}$ bin.}
\label{fig:EffArea} 
\end{figure}

Figure \ref{fig:EffArea} shows the result of the three generated speeds.
From Section \ref{sec:SSim}, $A_{\rm{gen}}^{\gamma}=1.33\,\rm{km}^{2}$ and 
$N_{\rm{gen}}^{\gamma}(\cos{\theta})=N_{\rm{gen}}^{\gamma}/20=5000$,
since
the generated flux is isotropic.  This is conservatively generalized to
any speed at the detector by treating the effective area as a step function, 
e.g. $A_{\rm{eff}}^{\gamma>10}=A_{\rm{eff}}^{\gamma=10}$, etc.  For $\beta<0.8$, the
effective area is set to zero.

\subsection{\label{sec:FinalLimitES}Limits on Isotropic Fluxes at the Earth's Surface}
The final limit on a flux with given mass and kinetic energy at the Earth's surface
($\Phi^{\gamma_{s},M}_{\mathrm{90\% C.L.}}$)
is calculated by scaling a reference flux with
the ratio of the Feldman-Cousins upper limit ($\mu_{\mathrm{90\%}}$) \cite{FeldCous}
to the expected number of signal events seen in the detector using the reference
flux.
The expected signal event number is found by going through each $\cos{\theta}$
bin and determining (1) what speed the monopole will have at the detector 
($\gamma_{d}$) based on
Section \ref{sec:AngAccEarth}
and (2) calculating the effective area for that speed and $\cos{\theta}$ bin 
based on Section \ref{sec:AngAccAnalysis}.
The final flux limit becomes:
\begin{eqnarray}
\Phi^{\gamma_{s},M}_{\mathrm{90\% C.L.}}=\frac{\mu_{\mathrm{90\%}}(N_{\rm{obs}}=0.0,N_{\rm{bkg}}=0.124)}{N_{\rm{sig}}(\Phi^{\gamma_{s},M}_{\rm{Ref}})}\Phi^{\gamma_{s},M}_{\rm{Ref}}
\\
N_{\rm{sig}}\approx T_{\rm{live}}\Phi^{\gamma_{s},M}_{\rm{Ref}} 2\pi \sum_{i=1}^{20}(\Delta\cos{\theta})_{i} A^{\gamma_{d}}_{\rm{eff}}(\cos{\theta}_{i})
\end{eqnarray}

$T_{\rm{live}}=2.06\times10^{7}\,$s is the total livetime of the 
analyzed data set, $N_{\rm{bkg}}=0.124$ is the final background expectation
from Table \ref{tab:Rates},
$\Delta\cos{\theta}=0.1$ is the width of the $\cos{\theta}$ bins, 
and the $2\pi$ arises from the azimuthal symmetry of the Earth.  
For most tested values of $\gamma_{s},M$, the final speed is the same
for all bins and the flux limit calculation returns the same answer as Table
\ref{tab:FinalLimits}.

\begin{figure}
\includegraphics[scale=0.4]{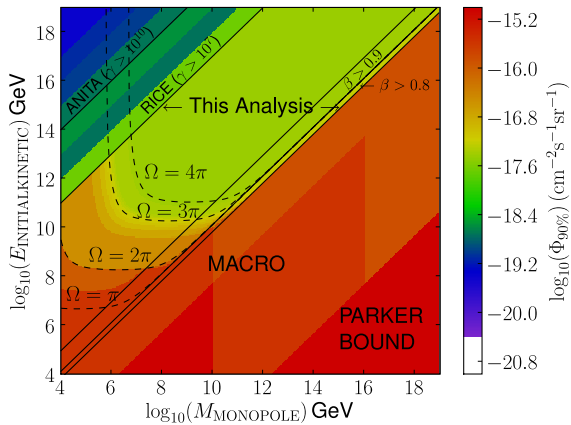}
\caption{Final flux limits (90\% C.L.) as function of monopole mass and kinetic energy at 
Earth's surface.  For relativistic mass and energies, only the most
restrictive limit is displayed.  Includes the Parker Bound \cite{PB} and results from 
MACRO \cite{MACRO}, RICE \cite{RICE}, and ANITA \cite{ANITA}.  For numerical values
of the final result for this analysis, see \cite{MyThesis}.}
\label{fig:finalresult} 
\end{figure}

To place this result in context, Fig. \ref{fig:finalresult}
displays the current best experimental flux limits over a wide range of mass and
kinetic energy values of magnetic monopoles.  
Below $\gamma=1.67$ the analysis does not apply as the 
monopoles fall below the Cherenkov threshold, while above $\gamma=10^{7}$, 
the radio neutrino detectors offer better sensitivity.

For the range of mass/kinetic energy pairs resulting
in $1.67<\gamma<10^{7}$, this analysis provides in general the best
flux limits to date.  
The exception occurs for the smallest masses and kinetic energies,
where attenuation in the Earth affects the signal acceptance.
To help guide the eye,
lines showing the angular acceptance solid angle $\Omega$
for the $\beta=0.9$
magnetic monopoles are included.  The solid angle is found by multiplying 2$\pi$
by the range of $\cos{\theta}$ for which the mass and energy
combination can reach the detector.  Hence the shape matches Fig.
\ref{fig:PSpace}.  As the solid angle approaches 2$\pi$, acceptance
below the horizon is lost and the limit becomes much weaker.

For the cases where $\gamma>10^{4}$, the flux limit from this analysis is 
conservative, as the monopole would have a large light contribution from
secondary cascades which are not yet included in the simulation.  These
will make the event brighter in the detector and increase the selection efficiency.

\section{\label{sec:Conc}Conclusion}
This analysis is the first search for magnetic monopoles using
the next generation of neutrino telescopes.  A final flux limit of 
$\Phi_{\mathrm{90\%C.L.}} = 3.38\e{-18}\fluxunits$ for $\beta\geq0.995$
is found.  For speeds down to $\beta=0.8$, the flux limit is slightly higher.
This applies to an isotropic flux
at the Earth's surface for relativistic magnetic monopoles with mass above 
$\sim10^{6}\,$GeV and energy above $\sim10^{10}\,$GeV 
(Fig. \ref{fig:finalresult}).  
Even with a single year of data
operating at $\sim20\,$\% of the final instrumented volume,
experimental flux limits are
achieved that are a factor of 4 below the current best constraints up to 
$\gamma\sim10^{7}$
and provide a good compliment to the more sensitive radio searches for
ultra-relativistic monopoles.  

This analysis does not follow IceCube's usual procedure of a blind analysis.
An original analysis performed on the data was done in
a blind fashion, with cuts being determined by simulation
datasets along with a 10\% 'burn' sample of experimental data.
It aimed to enhance the sensitivity to slower monopoles
by binning the data based on speed reconstruction.  Unblinding
revealed deficiencies in the background simulation to reproduce
the tails of the speed distribution where the slower signal
should be and allowed obvious background events into the final
sample.  After determining no monopole events were recorded, 
the analysis reported here is performed with cuts optimized on
improved simulation and not the experimental data.  The only changes 
involve a slight tightening of quality cuts motivated by the new 
simulation and abandoning the binning based on speed reconstruction.  For a full description 
of the original analysis, final event rejection, 
and motivation for changes, see \cite{MyThesis}.

Preliminary work on the 2008 data run shows
that the increased detector size and improvements to the analysis method 
provide a further factor of 
4 reduction in the sensitivity \cite{Jonas}.  
With more data and refined techniques,
IceCube and other neutrino telescopes will continue to prove valuable
in searches for
magnetic monopoles in the relativistic regime.

\begin{acknowledgments}
We acknowledge the support from the following agencies:
U.S. National Science Foundation-Office of Polar Programs,
U.S. National Science Foundation-Physics Division,
University of Wisconsin Alumni Research Foundation,
the Grid Laboratory Of Wisconsin (GLOW) grid infrastructure at the University of Wisconsin - Madison, the Open Science Grid (OSG) grid infrastructure;
U.S. Department of Energy, and National Energy Research Scientific Computing Center,
the Louisiana Optical Network Initiative (LONI) grid computing resources;
National Science and Engineering Research Council of Canada;
Swedish Research Council,
Swedish Polar Research Secretariat,
Swedish National Infrastructure for Computing (SNIC),
and Knut and Alice Wallenberg Foundation, Sweden;
German Ministry for Education and Research (BMBF),
Deutsche Forschungsgemeinschaft (DFG),
Research Department of Plasmas with Complex Interactions (Bochum), Germany;
Fund for Scientific Research (FNRS-FWO),
FWO Odysseus programme,
Flanders Institute to encourage scientific and technological research in industry (IWT),
Belgian Federal Science Policy Office (Belspo);
University of Oxford, United Kingdom;
Marsden Fund, New Zealand;
Australian Research Council;
Japan Society for Promotion of Science (JSPS);
the Swiss National Science Foundation (SNSF), Switzerland.
\end{acknowledgments}

\end{document}